\documentclass[12pt]{article}
\usepackage{graphics}
\usepackage{graphicx}
\DeclareGraphicsExtensions{.pdf}
\usepackage{float} 
\usepackage{amsmath}
\usepackage{slashed}
\usepackage{amsfonts}   
\usepackage{amssymb}

\usepackage{color}
\renewcommand\theequation{{\color{blue}\arabic{equation}}}
\begin{document}
\date{}
\title{{\bf{\Large Non-integrability for $ \mathcal{N}=1 $ SCFTs in $ 5d $  }}}
\author{
 {\bf {\normalsize Dibakar Roychowdhury}$
$\thanks{E-mail:  dibakarphys@gmail.com, dibakar.roychowdhury@ph.iitr.ac.in}}\\
 {\normalsize  Department of Physics, Indian Institute of Technology Roorkee,}\\
  {\normalsize Roorkee 247667, Uttarakhand, India}
\\[0.3cm]
}

\maketitle
\begin{abstract}
We explore the \emph{Liouvillian} non-integrability criteria for long quiver gauge theories those preserve $ \mathcal{N}=1 $ SUSY in $ 5d $.  We probe type IIB solutions with (an $ AdS_6 $ factor) semi-classical strings which capture the strong coupling dynamics of $ \mathcal{N}=1 $ SCFTs in $ 5d $. Our analysis reveals an underlying non-integrable structure within some sub-sector of these $ 5d $ SCFTs. To solidify our claim, we complement our analytic results through numerics. We estimate various chaos indicators for the phase space which confirm the onset of a \emph{chaotic} motion for these type IIB strings. 
\end{abstract}
\section{Overview and Motivation}
\subsection{Type IIB solutions with $ AdS_6 $ factor}
Among the plethora of examples of SCFTs those exist in diverse dimensions (and preserve different amount of SUSY), $ \mathcal{N}=1 $ SCFTs in five dimesions are of particular interest. Starting with the seminal work due to Seiberg \cite{Seiberg:1996bd} and thereby subsequently followed by authors in \cite{Brandhuber:1999np}-\cite{Passias:2012vp}, recently this particular class of SCFTs has attracted a lot of attention in the context of the holographic correspondence.

The bulk dual of these SCFTs is realised as a type IIB solution with an $ AdS_6 $ factor those preserving only half of the total thirty-two supersymmetries in $ 10d $. In the original construction of \cite{DHoker:2016ujz}-\cite{DHoker:2017zwj}, these type IIB solutions were realised as a warped product of the form $ AdS_6 \times S^2 \times \Sigma_{(2)} $ where $  \Sigma_{(2)}  $ is a two dimensional Riemann surface parametrised by means of some complex coordinates ($ z , \bar{z}$). Furthermore, in the original constructions of \cite{DHoker:2016ujz}-\cite{DHoker:2017zwj}, the warped factors of $ AdS_6 $ as well as $ S^2 $ line elements are defined in terms of the \emph{locally} holomorphic functions on the two dimensional Riemann surface ($ \Sigma_{(2)}  $).

Following these developments, several other aspects of $ \mathcal{N}=1 $ quiver gauge theories in $ 5d $ have been explored in the recent years. Let us review some of these results here. 

A large class of $ 5d $ SCFTs and the RG flow between these theories have been explored recently by authors in \cite{Fluder:2020pym}. Their analysis reveals an intriguing fact namely they show that the sphere free energy for these theories decreases as one RG flows from UV to IR. 

Codimension 2 (surface) defects in $ 5d $ SCFTs have been investigated by authors in \cite{Gutperle:2020rty} using ($ p , q $) five brane webs with D3 branes. In a related study, non local operators - for example Wilson loops in $ 5d $ SCFTs have been constructed in \cite{Uhlemann:2020bek}. Other than these, a series of papers \cite{Uhlemann:2019ypp}-\cite{Gutperle:2021nkl} have been put forward in the recent years those explore various other properties of these $ 5d $ SCFTs using type IIB solutions in $ 10d $.

In spite of these developments, several other field theoretic aspects of $ 5d $ SCFTs are yet to be addressed. One of these aspects include the possibility of finding integrability for these $ 5d $ fixed points. In a holographic set up, this translates into an equivalent question of showing integrability for the classical $ 2d $ world-sheet theory on $ AdS_6 \times S^2 \times \Sigma_{(2)} $. 

Showing classical integrability for $ 2d $ sigma models is in general is a difficult task as there is no general prescription for writing down the corresponding Lax pairs. An alternative approach might therefore be disproving the classical integrability for these type IIB strings on $ AdS_6 \times S^2 \times \Sigma_{(2)} $. For the purpose of the present paper, we choose to work with this second path where we probe the internal manifold of the full type IIB background by various wrapped string configurations. In a holographic framework, these solitons capture the dynamics of long/heavy single trace operators in the dual $ 5d $ SCFTs. The idea is to check the classical non-integrability for each of these configurations.

In this work, we adopt the recently proposed \emph{electrostatic} viewpoint \cite{Legramandi:2021uds}  of type IIB solutions with an $ AdS_6 $ factor\footnote{Typically, in an ``electrostatic description'' one classifies the quiver using the so called \emph{Rank} function ($ \mathcal{R}(\eta) $) which encodes all the information about the color and flavor nodes of the quiver. In electrostatic approach of \cite{Legramandi:2021uds}, one can embed the $ \mathcal{N}=1 $ quivers in a Hanany-Witten brane set up that comprises of NS5-D5-D7 brane intersections in $ 10d $. Here D5s correspond to the color nodes of the quiver whereas on the other hand, D7s are the flavor branes localized along the internal manifold of the type IIB background. For example, in this language $ \tilde{T}_{N_c,P} $ quivers are characterized by a linearly increasing Rank function ($ \mathcal{R}(\eta)=N_c \eta $) for $ 0 \leq \eta \leq (P-1) $ which is closed at $ \eta =P $ by placing flavor branes at $ \eta = P-1 $.}. Following this, we first review the basics of this electrostatic framework of type IIB solutions those preserve $ \mathcal{N}=1 $ SUSY. The full $ 10d $ solution may be expressed in terms of an $ AdS_6 $ factor together with some internal manifold ($ \mathcal{M}_4 $) that (contains a two sphere ($ S^2 $)) preserves the $ SU(2)_R $ symmetry of the dual SCFTs in $ 5d $. In our analysis, we probe this internal space with various semiclassical F1 strings. 

The type IIB background, that we choose to work with can be related to those obtained in \cite{Apruzzi:2018cvq} through S duality \cite{Legramandi:2021uds}. In the string frame, the type IIB background reads as\footnote{See Appendix B of \cite{Legramandi:2021uds} for an illuminating discussion on the mapping of type IIB solutions (\ref{e1})-(\ref{e8}) to the original construction by authors in \cite{DHoker:2016ujz}. This shows the equivalence between these two approaches.} \cite{Legramandi:2021uds}
\begin{eqnarray}
\label{e1}
ds^2_{IIB}&=&f_1 (\sigma , \eta) ds^2_{AdS_6}+ds^2_{\mathcal{M}_4}\\
&=& f_1 (\sigma , \eta) ds^2_{AdS_6}+f_2(\eta , \sigma)d\Omega_2 (\chi , \xi) +f_3 (\eta , \sigma)(d\sigma^2 +d\eta^2)\\
B_2 &=&f_4 (\sigma , \eta)\sin\chi d\chi \wedge d\xi ~~;~~\mathcal{C}_2 = f_5 (\sigma , \eta)\sin\chi d\chi \wedge d\xi\\
e^{-2\phi}&=&f_6(\sigma , \eta)~~;~~\mathcal{C}_0 = f_7 (\sigma , \eta)\\
f_1 &=&\frac{3\pi}{2}\left(\sigma^2 +\frac{3\dot{V}}{\partial^2_{\eta}V} \right)^{1/2}~;~f_2 = f_1\frac{\partial^2_{\eta}V \dot{V}}{3 \sigma \Delta}~;~ f_3 = f_1\frac{\partial^2_{\eta}V}{3\dot{V}}\\
f_4 &=&\frac{\pi}{2}\left(\eta -\frac{\dot{V}\partial_{\sigma}\partial_{\eta}V}{\Delta} \right) ~;~f_5 = \frac{\pi}{2}\left(V-\frac{\dot{V}}{\Delta}(\partial_{\eta}V(\partial_{\sigma}\partial_{\eta}V)-3\partial^2_{\eta}V \partial_{\sigma}V) \right)\\
f_6 &=& \frac{12\sigma \dot{V}\partial^2_{\eta}V \Delta}{(3\partial_{\sigma}V+\sigma \partial^2_{\eta}V)^2}~;~f_7 = 2 \left(\partial_{\eta}V+\frac{3\dot{V}\partial_{\sigma}\partial_{\eta}V}{(3\partial_{\sigma}V+\sigma \partial^2_{\eta}V)} \right) \\
\Delta &=& \frac{1}{\sigma}(2\dot{V}-\ddot{V})\partial^2_{\eta}V +\sigma (\partial_{\sigma}\partial_{\eta}V)^2~;~\dot{V}(\sigma , \eta)=\sigma \partial_{\sigma}V.
\label{e8}
\end{eqnarray}  

The BPS conditions yield the following PDE for the potential function $ V(\sigma , \eta) $
\begin{eqnarray}
\partial_{\sigma}(\sigma^2 \partial_{\sigma}V)+\sigma^2 \partial^2_{\eta}V=0
\end{eqnarray}
that is required to be solved with appropriate boundary conditions namely
\begin{eqnarray}
\hat{V} (\sigma \rightarrow \pm \infty , \eta ) =0~~;~~\mathcal{R} (\eta =0) =0=\mathcal{R} (\eta =P).
\end{eqnarray}

The potential function $ \hat{V} = \sigma V $ satisfies Laplace equation of electrostatics with the boundary conditions 
\begin{eqnarray}
\hat{V} (\sigma , \eta =0 ) =0=\hat{V} (\sigma , \eta =P )
\end{eqnarray}
where the range for holographic direction ($ \eta $) is bounded between $ 0 $ and $ P $. Given the above electrostatic equivalence, one can interpret $ \mathcal{R} (\eta)$ as the charge distribution (along the holographic axis ($ \eta $)) between two conducting planes placed at $ \eta =0 $ and $ \eta =P $ \cite{Legramandi:2021uds}. 

The Hanany-Witten set up corresponding to (\ref{e1})-(\ref{e8}) consists of an intersection of NS5-D5-D7 brane configuration in $ 10d $. Clearly, the corresponding $ \mathcal{N}=1 $ superconformal quiver must end at $ \eta =P $ which is achieved by placing flavour D7 branes at that point. 

In the present paper, we restrict ourselves to $ \sigma \sim 0$ plane while moving along the $ \eta $ direction of $ \mathcal{M}_4 $. Namely, we consider a legitimate expansion of the potential function $ \hat{V}(\sigma, \eta) $ near $ \sigma \sim 0 $ and estimate the corresponding metric functions $ f_i (\sigma , \eta) $.  Under such circumstances, one therefore expects a charge distribution between the conducting planes at $ \eta =0 $ and $ \eta=P $. This depends on the location of the flavour D7 branes along the $ \eta $ axis of $ \mathcal{M}_4 $. This is equivalent of saying that the corresponding rank function ($ \mathcal{R}(\eta) $) of the associated $ SU(N_c) $ color (gauge) group is piecewise linear in the interval $ 0 \leq \eta \leq P $.
\subsection{Summary of results}
Given the above set up (\ref{e1})-(\ref{e8}), below we summarize the key findings of the paper. We explore the strong coupling dynamics of $ \mathcal{N}=1 $ linear quivers by probing the type IIB geometry (\ref{e1})-(\ref{e8}) with various semiclassical F1 string configurations.\\\\
$ \bullet $ Our analysis reveals an intriguing fact - while certain long operators (at strong coupling) exhibit a simple set of dispersion relations, the $\mathcal{N}=1  $ superconformal fixed point (in $ 5d $) in general maintains some underlying non-integrable structure with it. We confirm this non-integrable structure using both analytic as well as numeric techniques.\\\\
$ \bullet $ The analytic technique that we implement in this paper is based on the rigorus mathematical formalism due to Kovacic\footnote{See Section \ref{3.1} for a brief discussion on the algorithm due to Kovacic.} \cite{kov1}-\cite{kov2} that has been applied (in order to unfold non-integrability for a wider class of supersymmetric gauge theories in diverse dimensoions) with a remarkable success in the recent years \cite{Basu:2011fw}-\cite{Rigatos:2020hlq}. 

Following the methodology as discussed in \cite{Basu:2011fw}-\cite{Stepanchuk:2012xi}, we propose a consistent $ 1d $ reduction of type IIB sigma model that fails to be compatible with the criteria set by Kovacic which therefore disproves the Liouvillian integrability of the sigma model in general. By virtue of the holographic correspondence, this translates into the simple fact of the non existence of the integrable structure for some specific (sub)sector of $ \mathcal{N}=1 $ SCFTs in $ 5d $.\\\\
$ \bullet $ In order to solidify our claim, we complement our analytic results through numerical studies where we estimate various chaos indicators \cite{Zayas:2010fs}-\cite{Basu:2011di} for the theory under consideration. Our analysis reveals that the non-integrability in $ \mathcal{N}=1 $ SCFTs triggers a \emph{chaotic} motion associated with the phase space trajectories of the type IIB (super)strings which in turn also destroys the so called KAM tori \cite{Basu:2011dg}-\cite{Basu:2012ae} of integrable string trajectories.\\\\
$ \bullet $ We now summarize our results by referring to the different Sections of the paper. In Section \ref{S2}, we explore certain class of long operators in $ \mathcal{N}=1 $ SCFTs those are dual to folded fundamental (F1) strings which are allowed to pass through \emph{localized} flavour D7 branes along the internal space of the full type IIB solution. Our analysis reveals a set of simple dispersion relations for these long operator states\footnote{See Appendix \ref{A} for a detailed discussion on the Liouvillian (non)integrable structure for this particular stringy configuration.}. 

However, the main analysis of our paper refers to Section \ref{S3}, where we focus on a particular winding string ansatz that wraps the isometry of the internal two sphere ($ S^2 \subset \mathcal{M}_4 $) and fluctuates along the rest of the directions of the internal manifold ($ \mathcal{M}_4 $).  We show that a consistent $ 1d $ reduction of the original sigma model fails to be compatible with the analytic integrability criteria set by the Kovacic's algorithm \cite{kov1}-\cite{kov2}. 

In Section \ref{S4}, we further explore on the nature of this non-integrable deformation at the level of the Hamiltonian dynamics. We estimate various chaos indicators for example, the Lyapunov exponent as well as the Poincare section which together confirm the onset of a chaotic dynamics for these type IIB strings. 

Finally, we put forward some future remarks and draw our conclusion in Section \ref{S5}.
\section{Spectrum of long operators in $ \mathcal{N}=1 $ SCFTs}
\label{S2}
We begin with the description of extended (as well as \emph{folded}) F1 string configurations those probe type IIB geometry (\ref{e1}) along the $ \eta $ axis. While extended along the $ \eta $ direction, these strings naturally meet stack of flavour D7 branes those are localized along $ \mathcal{M}_4 $. Our goal would be to explore the imprint of these flavour D7 branes on the associated spectrum of long operators pertaining to $ \mathcal{N}=1 $ superconformal quivers at strong coupling.

In the supergravity approximation, these long (single trace) operators are dual to folded F1 strings whose dynamics is encoded in the following sigma model action\footnote{we set, $ \alpha'= g_s=1 $.}
\begin{eqnarray}
S_P &=& \frac{1}{4\pi} \int d\tau d \tilde{\sigma}\mathcal{L}_P\\
\mathcal{L}_P &=&-G_{MN}\partial_{\tau}X^M \partial_{\tau}X^N + G_{MN}\partial_{\tilde{\sigma}}X^M \partial_{\tilde{\sigma}}X^N +2B_{MN}\partial_{\tau}X^M \partial_{\tilde{\sigma}}X^N
\label{e13}
\end{eqnarray}
where we restrict ourselves only to the metric as well as the NS-NS sector of the full type IIB solution (\ref{e1})-(\ref{e8}).

The above Lagrangian (\ref{e13}) is supplemented with the Virasoro constraints of the following form
\begin{eqnarray}
T_{\tau \tau}&=&T_{\tilde{\sigma}\tilde{\sigma}}\nonumber\\
&=&G_{MN}\partial_{\tau}X^M \partial_{\tau}X^N + G_{MN}\partial_{\tilde{\sigma}}X^M \partial_{\tilde{\sigma}}X^N =0,\\
T_{\tau \tilde{\sigma}}&=&G_{MN}\partial_{\tau}X^M \partial_{\tilde{\sigma}}X^N=0.
\end{eqnarray}
\subsection{Long strings}
To start with, we place F1 strings near the center ($ \rho \sim 0 $) of global $ AdS_6 $. In particular, we consider long folded string configurations those are extended through the flavour D7 branes while simultaneously wrapping the $ S^2 $ along its equatorial plane.

These strings are therefore described by an embedding of the following form,
\begin{eqnarray}
\label{e16}
t = \tau ~;~\eta = \eta (\tilde{\sigma})~;~\sigma = \sigma (\tilde{\sigma})~;~\chi = \frac{\pi}{2}~;~\xi = \ell \tilde{\sigma}.
\end{eqnarray}

Given the above configuration (\ref{e16}), these strings are naturally decoupled from the background NS-NS fluxes. The effects of incorporating NS-NS coupling will be discussed in the subsequent sections.

Below we summarize the set of equations that readily follows from (\ref{e13})
\begin{eqnarray}
\label{e17}
2f_3 \sigma'' &=& \partial_{\sigma}f_3 (\eta'^2 -\sigma'^2)-2\partial_{\eta}f_3 \eta' \sigma' +\partial_{\sigma}f_1 + \ell^2 \partial_{\sigma}f_2,\\
2f_3 \eta'' &=& \partial_{\eta}f_3 (\sigma'^2 -\eta'^2)-2\partial_{\sigma}f_3 \eta' \sigma' +\partial_{\eta}f_1 + \ell^2 \partial_{\eta}f_2
\label{e18}
\end{eqnarray}
where the prime corresponds to derivative with respect to $ \tilde{\sigma} $.

The above set of equations (\ref{e17})-(\ref{e18}) are supplemented with Virasoro constraints of the following form
\begin{eqnarray}
\label{e19}
T_{\tau \tau}&=&T_{\tilde{\sigma}\tilde{\sigma}}\nonumber\\
&=&-f_1 + \ell^2 f_2 +f_3 (\eta'^2 + \sigma'^2) =0,\\
T_{\tau \tilde{\sigma}}&=&0.
\end{eqnarray}

Below we explore the above set of equations (\ref{e17})-(\ref{e19}) for different choices of the $ \mathcal{N}=1 $ linear quivers those were proposed recently in \cite{Legramandi:2021uds}.
\subsubsection{$ \tilde{T}_{N_c,P} $ quivers}
The first example we consider is that of single kink quivers ($ \tilde{T}_{N_c,P} $) those are closed at $ \eta =P $ by placing flavour D7 branes at $ \eta = P-1 $.

The corresponding rank function is given by
\begin{eqnarray}
\label{e21}
\mathcal{R}(\eta)= \begin{cases}
      N_c \eta &  0 \leq \eta \leq (P-1)\\
      N_c (P-1)(P- \eta) & (P-1)\leq \eta \leq P.
    \end{cases} 
\end{eqnarray}

In the supergravity approximation, the associated potential function reads as \cite{Legramandi:2021uds}
\begin{eqnarray}
\hat{V}(\sigma \sim 0 , \eta)\sim\frac{\eta N_c  P \log 2}{\pi }-\frac{\pi  \eta  \left(\eta ^2+1\right) N_c}{24 P}-\frac{\sigma \eta  N_c}{2}    +\mathcal{O}(\sigma^2 /P)
\label{e22}
\end{eqnarray}
where the above expansion (\ref{e22}) is valid in the regime where $ \sigma $ is finite and $ P \gg 1 $.

On the other hand, an expansion near $ |\sigma| \rightarrow \infty $ reveals a potential function of the following form\footnote{See Appendix \ref{B} for a detailed discussion on the large $ \sigma $ limit.}
\begin{eqnarray}
\label{e23}
\hat{V}(\sigma \rightarrow \infty , \eta)\sim \frac{P^3 N_c }{\pi ^3}e^{-\frac{\pi  \sigma }{P}} \sin \left(\frac{\pi }{P}\right) \sin \left(\frac{\pi  \eta }{P}\right).
\end{eqnarray}

Below, we explore the regime $ \sigma  \sim 0 $. The corresponding metric functions ($ f_i (\sigma , \eta) $) read as
\begin{eqnarray}
f_1 (\sigma \sim 0, \eta)&\sim &\frac{3\pi}{2}   \sqrt{-\frac{\eta ^2}{2}+\frac{12 P^2 \log 2}{\pi ^2}-\frac{1}{2}}+\mathcal{O}(\sigma^2) ~,\\
\label{e25}
f_2 (\sigma \sim 0, \eta)&\sim &\frac{-3 \left(\pi ^2 \eta ^2 \left(24 P^2 \log 2-\pi ^2 \left(\eta ^2+1\right)\right)^{3/2}\right)}{\sqrt{2} \left(\pi ^4 \left(9 \eta ^4+12 \eta ^2-1\right)-576 P^4 \log ^22+48 \pi ^2 \left(1-6 \eta ^2\right) P^2 \log 2\right)}\\
f_3 (\sigma \sim 0, \eta)&\sim &\frac{3 \pi ^2}{\sqrt{48 P^2 \log 2-2 \pi ^2 \left(\eta ^2+1\right)}}+\mathcal{O}(\sigma^2)
\end{eqnarray}
which clearly reveals that $ \partial_{\sigma}f_i (\sigma \sim 0, \eta)\sim 0 $. Therefore, $ \sigma = \sigma' =\sigma'' =0 $ is a natural solution of (\ref{e17}).

Using the above as the primary input, we compute the energy ($ E_{sk} $) associated with the folded string configuration which is dual to the conformal dimension associated with long operators in $ \mathcal{N}=1 $ single kink quivers at strong coupling
\begin{eqnarray}
\label{e27}
\hat{\Delta}_{sk}\sim E_{sk}\sim \frac{1}{\pi}\int_0^{P}f_1 \frac{d\eta}{\eta'}\sim \int_0^{P}d\eta \frac{f_1}{\sqrt{\Lambda_{sk}}}
\end{eqnarray}
where, we set the winding number $ \ell =1 $ together with
\begin{eqnarray}
\label{e28}
\Lambda_{sk}(\eta , P)&\sim & P^2 \log (4096)-\frac{1}{2} \pi ^2 \left(3 \eta ^2+1\right)+\frac{7 \pi ^4 \eta ^4}{P^2 \log (4096)}+\cdots \\
f_1 (\eta , P)& \sim & 3 P \sqrt{\log 8}-\frac{\pi ^2 \left(\eta ^2+1\right) \sqrt{\frac{3}{\log 2}}}{16 P}+\cdots
\label{e29}
\end{eqnarray}

In view of (\ref{e28})-(\ref{e29}), we see that the integral (\ref{e27}) yields a finite answer
\begin{eqnarray}
\label{e30}
\hat{\Delta}_{sk}|_{P\gg 1}\sim \frac{3 P}{2}\sim \frac{3}{2N_c}Q_{D7}
\end{eqnarray}
which shows that the dimension of the dual operator grows linearly with the size of the quiver. Here, $ Q_{D7}=P N_c $ is the Page charge associated with flavour D7 branes \cite{Legramandi:2021uds}.

A simple interpretation of the above result (\ref{e30}) comes from a closer inspection of the sigma model potential for F1 strings near flavour D7 branes. This readily follows from the Lagrangian density (\ref{e13})
\begin{eqnarray}
V_{eff}(\eta , P)\sim 3 P \sqrt{\log 8}+\frac{\pi ^2 \left(3 \eta ^2-1\right) \sqrt{\frac{3}{\log 2}}}{16 P}+\cdots
\end{eqnarray}
which shows that the potential is regular across the location of D7 branes. In other words, the F1 string can smoothly pass through flavour D7 branes without exhibiting any divergences in the spectrum.
\subsubsection{$ +_{P, N_c} $ quivers}
\label{2.1.2}
The second example we consider is that of a linear quiver ($ +_{P, N_c} $) with a plateau region which corresponds to placing stack of flavour D7 branes at $ \eta =1 $ and $ \eta =P-1 $. As we shall see shortly that these quivers seek special attention around the position $ \eta \sim 1 $.

The corresponding rank function reads as
\begin{eqnarray}
\label{e32}
\mathcal{R}(\eta)= \begin{cases}
      N_c \eta &  0 \leq \eta \leq 1\\
      N_c  &1\leq \eta \leq (P-1)\\
      N_c (P-\eta) &(P-1)\leq \eta \leq P.
    \end{cases} 
\end{eqnarray}

The associated potential function when expanded near $ \sigma \sim 0 $ reads as,
\begin{eqnarray}
\label{e33}
\frac{\hat{V}(\sigma , \eta)}{(\frac{N_c}{4 \pi})}\sim \eta  (6+4\log 2)-4 \eta  \log \left(\frac{\pi }{P}\right)-2\eta \log |1 - \eta^2 | -(1+ \eta^2 - \sigma^2) \log \left| \frac{\eta +1}{\eta -1}\right| 
\end{eqnarray}
where we have taken into account the large $ P(\gg 1) $ limit.

A careful analysis reveals that the potential function (\ref{e33})
\begin{eqnarray}
\frac{\hat{V}(\sigma \sim 0 , \eta \sim 1)}{(\frac{N_c}{4 \pi})}\sim 6-4  \log \left(\frac{\pi }{P}\right)+\mathcal{O}(\eta -1)
\end{eqnarray}
is indeed \emph{regular} across the location of flavour D7 branes at $ \eta =1 $. As a result, one should expect that the corresponding metric functions ($ f_i(\sigma , \eta) $) to be regular across $ \eta \sim 1 $.

In order to estimate the $ \eta $ integral as before, we therefore divide the entire domain of definition into three regions namely (I) Region I ($ 0 \leq \eta \leq 1-\delta $), (II) Region II ($ 1-\delta \leq \eta \leq 1+\delta $) and (III) Region III ($ 1+\delta \leq \eta \leq P$) where the parameter $ \delta $ is set to be zero towards the end of the calculation.

After a careful analysis, the metric functions corresponding Region I turn out to be
\begin{eqnarray}
f_1 (\sigma \sim 0, \eta <1)& \sim & \frac{3}{2} \pi  \sqrt{-\frac{\eta ^2}{2}-3 \log \left(\frac{\pi }{P}\right)+3+\log 8},\\
f_2 (\sigma \sim 0, \eta <1)& \sim & -\frac{\pi  \eta ^2 \left(-\eta ^2-6 \log \left(\frac{\pi }{P}\right)+6+\log (64)\right)^{3/2}}{3 \left(\sqrt{2} \left(\eta ^4+8 \eta ^2 \left(\log \left(\frac{\pi }{2 P}\right)-1\right)-4 \left(\log \left(\frac{\pi }{2 P}\right)-1\right)^2\right)\right)},\\
f_3 (\sigma \sim 0, \eta <1)& \sim & -\frac{3 \left(\pi  \sqrt{-\frac{\eta ^2}{2}-3 \log \left(\frac{\pi }{P}\right)+3+\log (8)}\right)}{\eta ^2+6 \log \left(\frac{\pi }{P}\right)-3 (2+\log (4))}.
\end{eqnarray}

Let us now explore spacetime solutions in Region II  which reveals
\begin{eqnarray}
f_1 (\sigma \sim 0, \eta \sim 1)& \sim &\frac{3\sqrt{3}\pi}{\sqrt{2}} k_c \sqrt{\left|2 \log \left(\frac{\pi }{P}\right)-3\right|},\\
f_2 (\sigma \sim 0, \eta \sim 1)& \sim &\frac{1}{9}f_1 (\sigma \sim 0, \eta \sim 1),\\
f_3 (\sigma \sim 0, \eta \sim 1)& \sim &\frac{9\pi^2}{4}f^{-1}_1 (\sigma \sim 0, \eta \sim 1),
\end{eqnarray}
where the ratio, $ \frac{\delta}{\sigma} =k_c$ is kept fixed in the limit $ \delta \rightarrow 0 $, $ \sigma \rightarrow 0 $.

Finally, we note down metric functions corresponding to Region III. These solutions are a bit lengthy which are therefore summarized in the Appendix \ref{C}.

Combining all these pieces together, the conformal dimension of the dual operator can be computed by performing the $ \eta $ integral piece wise
\begin{eqnarray}
\label{e41}
\hat{\Delta}_{qp}\sim  \frac{1}{\pi}\int_0^{1-\delta}f_1 \frac{d\eta}{\eta'} +\frac{1}{\pi}\int_{1-\delta}^{1+\delta}f_1 \frac{d\eta}{\eta'}+\frac{1}{\pi}\int_{1+\delta}^{P}f_1 \frac{d\eta}{\eta'}.
\end{eqnarray}

The first two terms on the R.H.S. of (\ref{e41}) are comparatively easy to evaluate. Notice that in each of these integrals $ \eta' $ can be substituted using the constraint (\ref{e19}). Computing the first two integrals in the holographic limit ($ \frac{\pi}{P} \ll 1 $) we find,
\begin{eqnarray}
\hat{\Delta}_{qp}&\sim & \frac{\zeta (P)}{2N_c}Q_{D7}~;~Q_{D7}=2N_c\\
\zeta (P)&= &\frac{3}{2}+\frac{1}{\pi}\int_{1}^{P}f_1 \frac{d\eta}{\eta'}
\end{eqnarray}
while the remaining integral is indeed difficult to evaluate for the entire range $ 1\leq \eta \leq P $. However, it is noteworthy to mention that in the holographic limit ($ P \gg 1 $) the dominant contribution to $ \zeta (P) $ comes from this remaining integral.

One can perform numerical integration which reveals the following set of data
\begin{eqnarray}
\label{e44}
\frac{1}{\pi}\int_{1}^{P}f_1 \frac{d\eta}{\eta'}= \begin{cases}
      156.603 &  P=100\\
      235.671  & P=150\\
      314.739 &P=200.
    \end{cases} 
\end{eqnarray}
From the above set of data, it is clear that for an increase $ \Delta P =50 $ there is an uniform increase ($ \sim 79 $) in the corresponding value of the integral (\ref{e44}). In other words, the function $ \zeta(P) $ increases linearly with $ P $ with a slope $ \sim \frac{3}{2} $. Therefore the energy of the quiver is roughly proportional to the size of the quiver
\begin{eqnarray}
\label{e45}
\hat{\Delta}_{qp}|_{P\gg 1} \sim \frac{3}{2}P .
\end{eqnarray}
This is precisely what we have seen in the case of single kink quivers (\ref{e30}). 
\subsection{Adding R charge}
We now generalize the previous analysis in the presence of non-zero R-charge ($ J $) which is the Cartan of the $ SU(2)_R $ symmetry of the internal $ S^2 $. In the dual stringy picture this corresponds to rotation of the string along the isometry direction of $ S^2 $.

The ansatz that we propose is of the form,
\begin{eqnarray}
t = \tau ~;~\eta = \eta (\tilde{\sigma})~;~\chi = \frac{\pi}{2}~;~\xi = \omega \tau
\end{eqnarray}
where, to begin with we set, $ \sigma = \sigma' =\sigma'' =0 $ as this turns out to be a solution of the resulting equations of motion. This is the simplest configuration that one could imagine where the string is stretched along the holographic ($ \eta $) axis while also rotating along $ \xi $.

The corresponding Lagrangian density is given by\footnote{In the subsequent analysis we set $ \omega =1 $ without any loss of generality.}
\begin{eqnarray}
\mathcal{L}_P=f_1 -\omega^2 f_2 + f_3 \eta'^2
\end{eqnarray}
which is supplemented with the Virasoro constraints of the following form
\begin{eqnarray}
T_{\tau \tau}&=&T_{\tilde{\sigma}\tilde{\sigma}}\nonumber\\
&=&-f_1 + \omega^2 f_2 +f_3 \eta'^2  =0,\\
T_{\tau \tilde{\sigma}}&=&0.
\end{eqnarray}

Given the above set up, below we estimate the energy ($ \Delta $) as well as the R- charge ($ J $) associated with long operators corresponding to each of the above quivers.
\subsubsection{$ \tilde{T}_{N_c,P} $ quivers}
Given the single kink quivers as depicted in (\ref{e21}), the energy ($ \hat{\Delta}_{sk} $) associated with the dual operator remains the same as in (\ref{e30}).

On the other hand, the R-charge corresponding to these dual operators is given by,
\begin{eqnarray}
\label{e49}
J=-\frac{1}{4 \pi}\int_0^{2\pi}d\tilde{\sigma}\frac{\delta \mathcal{L}_P}{\delta \dot{\xi}}=\frac{1}{\pi}\int_0^{P}f_2 \frac{d\eta}{\eta'}=\int_0^{P}d\eta \frac{f_2}{\sqrt{\Lambda_{sk}}}
\end{eqnarray}
where, in the holographic limit the denominator is given by an expansion of the form (\ref{e28}).

The numerator (\ref{e25}), on the other hand, can be expanded as
\begin{eqnarray}
\label{e50}
f_2 (\sigma \sim 0, \eta)|_{P\gg 1}\sim \frac{\pi ^2 \eta ^2 \sqrt{\frac{3}{\log 2}}}{4 P}+\mathcal{O}(1/P^3).
\end{eqnarray}

Substituting (\ref{e50}) into (\ref{e49}) we finally obtain
\begin{eqnarray}
\label{e51}
J \sim \frac{\pi ^2 P}{24 \log 2}.
\end{eqnarray}

Combining (\ref{e30}) with (\ref{e51}), we therefore conclude
\begin{eqnarray}
\hat{\Delta}_{sk}\sim \frac{36\log 2}{\pi^2} J \sim 2 J.
\end{eqnarray}
\subsubsection{$ +_{P, N_c} $ quivers}
A similar calculation as before reveals,
\begin{eqnarray}
\label{e54}
J \sim  \frac{1}{\pi}\int_0^{1-\delta}f_2 \frac{d\eta}{\eta'} +\frac{1}{9\pi}\int_{1-\delta}^{1+\delta}f_1 \frac{d\eta}{\eta'}+\frac{1}{\pi}\int_{1+\delta}^{P}f_2 \frac{d\eta}{\eta'}.
\end{eqnarray}

Like before, the dominant contribution to (\ref{e54}) comes from the integral in the range $ 1 \leq \eta \leq P $. We evaluate this integral numerically for three different choices of $ P $
\begin{eqnarray}
\label{e55}
\frac{1}{\pi}\int_{1}^{P}f_2 \frac{d\eta}{\eta'}= \begin{cases}
      15.7793 &  P=100\\
      23.7034  & P=150\\
      31.6258 &P=200
    \end{cases} 
\end{eqnarray}
which reveals that with an increase $ \Delta P =50 $, the value of $ J $ increases by an amount $ \sim 8 $.

Therefore, we propose that in the large $ P(\gg 1) $ limit
\begin{eqnarray}
\label{e56}
J \sim \frac{4}{25}P.
\end{eqnarray}

Combining (\ref{e56}) with (\ref{e45}) we find
\begin{eqnarray}
\hat{\Delta}_{qp}\sim \frac{75}{8}J \sim  9 J.
\end{eqnarray}
\subsection{Adding NS-NS flux}
We now generalize our previous analysis by coupling the sigma model with background NS-NS fluxes. Like before, we restrict ourselves to the $ \sigma =0 $ plane and propose an embedding of the following form
\begin{eqnarray}
\label{e58}
t = \tau ~;~\eta = \eta (\tilde{\sigma})~;~\chi = \chi (\tilde{\sigma})~;~\xi = \tau
\end{eqnarray}
which corresponds to extended F1 string configurations those are stretched simultaneously along $ \eta $ and $ \chi $ direction. Our purpose would be to check whether the above configuration (\ref{e58}) allows a sustainable configuration in the large $ P(\gg 1) $ limit.

The corresponding Lagrangian density turns out to be
\begin{eqnarray}
\label{e59}
\mathcal{L}_P=f_1 - f_2 \sin^2\chi +f_2 \chi'^2 + f_3 \eta'^2 -2f_4 \sin\chi \chi'.
\end{eqnarray}

Below, we note down the $ \chi $ equation of motion that readily follows from (\ref{e59})
\begin{eqnarray}
\label{e60}
f_2 \chi'' =-\frac{f_4}{2}\sin2\chi -\partial_{\eta}f_2 \eta' \chi' +\partial_{\eta}f_4 \sin\chi \eta'.
\end{eqnarray}

The corresponding Virasoro constraint reads as,
\begin{eqnarray}
\label{e61}
-f_1 +f_2 \chi'^2 +f_2 \sin^2 \chi +f_3 \eta'^2 \approx 0.
\end{eqnarray}

Below we investigate each of the above equations (\ref{e60})-(\ref{e61}) considering both the examples of single kink as well as quivers with plateau.
\subsubsection{$ \tilde{T}_{N_c,P} $ quivers}
Let us explore (\ref{e60}) in the holographic limit $ P\gg 1 $. An expansion of $ f_4(\sigma , \eta) $ in the holographic limit reveals
\begin{eqnarray}
f_4 (\sigma \sim 0, \eta)|_{P \gg 1}\sim \frac{\pi ^3 \eta ^3}{2P^2 \log 2}+\mathcal{O}(1/P^4)
\end{eqnarray}
which together with (\ref{e50}) yields an equation of the form
\begin{eqnarray}
\frac{\chi''}{\chi'}+\frac{2\eta'}{\eta}\approx 0
\end{eqnarray}
that has a simple solution of the form
\begin{eqnarray}
\label{e64}
\chi' (\tilde{\sigma})\sim \frac{\mathcal{C}}{\eta^2} +\mathcal{O}(1/P)
\end{eqnarray}
where $ \mathcal{C} $ is a constant of integration. 

Notice that, the function (\ref{e64}) is singular as $ \eta $ approaches zero. Therefore, in order for the Lagrangian (\ref{e59}) to be well defined throughout the range $ 0 \leq \eta \leq P $ one must set $ \mathcal{C}=0 $. In other words,  $ \chi (\tilde{\sigma})\sim $ constant in the strict holographic limit which naturally decouples the F1 string from the background NS-NS fluxes. Therefore, our findings essentially boil down to those in the absence of NS-NS fluxes.

To see this explicitly, we plug (\ref{e64}) into (\ref{e61}) which reveals
\begin{eqnarray}
\label{e65}
\eta' (\tilde{\sigma})\sim \frac{2\sqrt{3}\sqrt{\log2}}{\pi} ~P+\cdots
\end{eqnarray}

Using (\ref{e65}), the spectrum of long operators finally meet our expectation namely
\begin{eqnarray}
\hat{\Delta}_{sk}\sim \frac{3}{2}P.
\end{eqnarray}

A similar calculation for the R-charge can be carried out in parallel which yields same answer as in (\ref{e51}).
\subsubsection{$ +_{P, N_c} $ quivers}
We adopt similar methodology for quivers with a plateau where we estimate the metric functions ($ f_i(\sigma \sim 0 , \eta) $) for the range $ 1\lesssim \eta \leq P $ as this produces dominant contributions to the integrals in the holographic limit.  

Below, we summarize the metric functions $ f_i(\sigma \sim 0 , \eta) $ in the large $ P (\gg \pi)$ limit
\begin{eqnarray}
\label{e67}
f_1 (\sigma \sim 0 , \eta)& \sim &3 \sqrt{\frac{3}{2}} \pi  \sqrt{-\frac{\eta  \log \left(\frac{\pi }{P}\right)}{\log \left(\frac{\eta +1}{\eta -1}\right)}},\\
f_2 (\sigma \sim 0 , \eta)& \sim &\frac{-9\pi^2  \eta ^2 \log \left(\frac{\pi }{P}\right)}{2f_1 \left(-2 \log \left(\eta ^2-1\right)+\eta  \log \left(\frac{\eta +1}{\eta -1}\right)-2 \log \left(\frac{\pi }{P}\right)+4+\log 16\right)},\\
f_3 (\sigma \sim 0 , \eta)& \sim &\frac{9 \pi^2}{4 f_1},\\
f_4(\sigma \sim 0, \eta) &\sim &\frac{\pi  \left(\left(5 \eta ^2+1\right) \log \left(\frac{\eta +1}{\eta -1}\right)-2 \eta \right)}{4 \left(-2 \log \left(\eta ^2-1\right)+\eta  \log \left(\frac{\eta +1}{\eta -1}\right)-2 \log \left(\frac{\pi }{P}\right)+4+\log 16\right)}.
\label{e70}
\end{eqnarray}

Using (\ref{e67})-(\ref{e70}), it is now straightforward to show that in the strict holographic limit ($ \frac{\pi}{P} \sim 0 $) one has $ f_4 \sim \partial_{\eta}f_4 \sim 0$. On the other hand, a straightforward computation reveals
\begin{eqnarray}
\frac{\partial_{\eta}f_2}{f_2}\Big|_{P \rightarrow \infty} \sim \frac{1}{\eta}
\end{eqnarray}
which by virtue of (\ref{e60}) and following our previous discussion yields a solution of the form
\begin{eqnarray}
\chi' (\tilde{\sigma}) \sim \mathcal{O}(1/P)\sim 0.
\end{eqnarray}

Therefore, to summarise, we conclude that in the strict holographic limit, the NS-NS flux does not affect the spectrum of long operators in $ \mathcal{N}=1 $ SCFTs.
\section{Liouvillian non-integrability}
\label{S3}
\subsection{The algorithm}
\label{3.1}
The strong coupling behaviour of both $ \tilde{T}_{N_c,P} $ as well as $ +_{P, N_c} $ quivers are encoded in the dynamics associated with type IIB strings those are described by the classical sigma model Lagrangian of the form (\ref{e13}). Therefore, one way to prove or disprove the integrability for these long quivers is to adopt some algorithm that classifies the underlying integrable structure associated with the phase space dynamics of these (semi)classical strings. Below, we elaborate more on this algorithm that drives the rest of our analysis. 

To prove the integrability of the $ 2d $ sigma model (\ref{e13}) one needs to find the corresponding Lax pairs that reproduce the dynamics of the string in a consistent manner. This is indeed a non-trivial task. Therefore, instead of finding the Lax pairs, one should look for an alternative that disproves integrability for some particular embedding of these (semi)classical strings. This algorithm is named after Kovacic \cite{kov1}-\cite{kov2} that tells us some set of rules to classify the phase space dynamics and the associated integrable structure. 

To apply the machinery due to Kovacic, the first step is to consider a consistent $ 1d $ truncation of the original sigma model (\ref{e13}) and study the resulting dynamics. The truncation usually results in a set of coupled PDEs those can be solved either numerically or analytically. Following the algorithm closely, one can in fact reduce these PDEs into a linear second order differential equation called the normal variational equation\footnote{Here, by ``dot" we mean derivative with respect to $ \tau $.} (NVE)
\begin{eqnarray}
\label{e73}
\ddot{y}(\tau)+\mathcal{B}(\tau)\dot{y}(\tau)+\mathcal{A}(\tau)y(\tau)=0
\end{eqnarray}
where $ \mathcal{A} $ and $ \mathcal{B} $ are (complex) rational functions.

Given the NVE (\ref{e73}), one concludes that the classical phase of the string soliton is \emph{Liouvillian} integrable if the corresponding solution can be expressed in terms of simple algebraic polynomials, harmonic functions, exponential or logarithmic functions - collectively known as Liouvillian form of solutions \cite{Basu:2011fw}-\cite{Nunez:2018qcj}.

In his pioneering work \cite{kov1}, Kovacic has clearly stated about the \emph{necessary} (but not sufficient) conditions for NVE to admit Liouvillian form of solutions and prescribed an algorithm to construct such solutions based on the notion of general $ SL(2,C) $ group of invariance of (\ref{e73}). Let us elaborate on these conditions and summarize all the key features.

To understand these conditions properly, one needs to convert the NVE (\ref{e73}) into the familiar Schroedinger form
\begin{eqnarray}
\label{e74}
\dot{\omega} (\tau)+\omega^{2}(\tau)=V(\tau)=\frac{2\dot{\mathcal{B}}+\mathcal{B}^2 -4 \mathcal{A}}{4}
\end{eqnarray}
by redefining the original variable as
\begin{eqnarray}
y(\tau)=e^{\int (\omega(\tau)-\frac{\mathcal{B}(\tau)}{2})d\tau}.
\end{eqnarray}

The NVE (\ref{e73}) allows a Liouvillian form of solution if the function $ \omega(\tau) $ turns out to be a simple algebraic polynomial of degree 1, 2, 4, 6 or 12 \cite{kov2}. In his original work \cite{kov1}, Kovacic clearly mentioned about those necessary conditions \cite{Nunez:2018qcj} that the potential function ($ V(\tau) $) must satisfy for the algorithm to be applicable in the first place.

These conditions essentially talk about the pole structure of $ V(\tau) $ both for finite as well as large values of $ \tau $. They are summarized quite nicely in the Appendix of \cite{Nunez:2018qcj}.  Once one of these criteria are satisfied then the algorithm can be applied to categorize the solutions of (\ref{e74}) into one of the above polynomials. On the other hand, if none of these (minimal) criteria are satisfied, then the analytic solution of (\ref{e73}) turns out to be non-Liouvillian and hence the corresponding phase space dynamics is non-integrable.

Below, we elaborate on this taking specific examples of $ \mathcal{N}=1 $ linear quivers in $ 5d $.
\subsection{$1d$ reduction}
We begin by considering a consistent $ 1d $ reduction of the original sigma model (\ref{e13}). To this end, we propose an embedding of the following form
\begin{eqnarray}
\label{e76}
t = \tau ~;~\eta = \eta (\tau)~;~ \chi = \chi (\tau)~;~ \xi = \ell \tilde{\sigma}
\end{eqnarray}
where, we place the string soliton at the center of $ AdS_6 $ and restrict ourselves to the $ \sigma =0 $ plane of the internal manifold $ \mathcal{M}_4 $. For simplicity, we consider the coupling of the string with the metric and the background NS-NS flux.

Using (\ref{e76}), the Lagrangian density for the reduced model turns out to be
\begin{eqnarray}
\label{e77}
\mathcal{L}^{(1d)}_P = f_1 \dot{t}^2 -f_2 \dot{\chi}^2 -f_3 \dot{\eta}^2 +\ell^2 f_2 \sin^2\chi +2\ell f_4 \sin\chi \dot{\chi}.
\end{eqnarray}

Below, we note down the conjugate momenta
\begin{eqnarray}
p_t&=&E=2 \dot{t}f_1,\\
p_{\chi}&=&-2f_2 \dot{\chi}+2 \ell f_4 \sin\chi ,\\
p_{\eta}&=&-2f_3 \dot{\eta},
\end{eqnarray}
that lead to the Hamiltonian density of the following form
\begin{eqnarray}
\label{E81}
-\mathcal{H}^{(1d)}=-\frac{E^2}{4f_1}+\frac{p^2_{\eta}}{4f_3}+\frac{1}{4f_2}(p_{\chi}-2\ell f_4 \sin\chi)^2+\ell^2 f_2 \sin^2\chi.
\end{eqnarray}

Below, we note down the equations of motion that readily follow form (\ref{e77})
\begin{eqnarray}
\label{e81}
2f_3 \ddot{\eta}&=&\partial_{\eta}f_2 (\dot{\chi}^2 - \ell^2 \sin^2\chi)-\partial_{\eta}f_3 \dot{\eta}^2 -2 \ell \partial_{\eta}f_4 \sin\chi \dot{\chi}-\partial_{\eta}f_1 , \\
-f_2 \ddot{\chi}&=&\partial_{\eta}f_2 \dot{\eta}\dot{\chi}+\ell^2 f_2 \sin\chi \cos\chi -\ell \partial_{\eta}f_4 \dot{\eta}\sin\chi.
\label{e82}
\end{eqnarray}

Finally, we note down the Virasoro constraints associated with the sigma model
\begin{eqnarray}
\label{e84}
-\mathcal{H}^{(1d)}&=&T_{\tau \tau}\nonumber\\
&=&-f_1 +f_3 \dot{\eta}^2 +f_2 \dot{\chi}^2 +\ell^2 f_2 \sin^2\chi =0, \\
T_{\tau \tilde{\sigma}}&=&0.
\end{eqnarray}

A straightforward computation further reveals
\begin{eqnarray}
\label{e85}
\partial_{\tau}T_{\tau \tau}=-2\partial_{\eta}f_1 \dot{\eta}=0,
\end{eqnarray}
which therefore demands that for (\ref{e76}) to be a consistent embedding one must set, $ \eta = \eta_s= $ constant. In other words, the consistency requirement of the Virasoro constraint (\ref{e85}) confines the stringy dynamics over the submanifold $ R \times S^2 $. This further restricts the phase space dynamics of the string soliton to the two dimensional subspace (of the full four dimensional phase space) characterized by $\lbrace p_{\eta}=0, \eta = \eta_s \rbrace$.

As we shall see, the reduced sigma model (\ref{e77}) allows two possible forms of NVEs. The solitonic configuration fails to be Liouvillian integrable if one of these NVEs does not meet the Kovacic's criteria those were elaborated above.
\subsection{NVEs}
\subsubsection{Case I}
\label{3.3.1}
In order to arrive at the corresponding NVEs, we choose to work with the invariant plane $ \lbrace p_{\chi}=0, \chi =0 \rbrace$ \cite{Basu:2011fw}-\cite{Nunez:2018qcj} in the phase space and consider fluctuations ($ y(\tau) $) normal to this plane. These fluctuations result in what we identify as NVEs those were mentioned previously in (\ref{e73}). Notice that the above subspace naturally solves (\ref{e82}) as $ \chi = \dot{\chi}=\ddot{\chi}=0 $.

It is indeed quite straightforward to figure out this constant ($ \eta_s $) for the given choice of phase space variables. A straightforward substitution into (\ref{e81}) reveals the condition
\begin{eqnarray}
\label{e86}
\partial_{\eta}f_1 (\sigma \sim 0 , \eta =\eta_s)\sim 0.
\end{eqnarray}

For $ \tilde{T}_{N_c,P} $ quivers (\ref{e21}), the condition (\ref{e86}) reveals
\begin{eqnarray}
\partial_{\eta}f_1 (\sigma \sim 0 , \eta =\eta_s)\Big |_{P \gg 1}\sim -\frac{\pi ^2 \eta_s  \sqrt{\frac{3}{\log 2}}}{8 P} -\frac{\pi ^4 \left(\eta_s ^2+1\right) \eta_s }{128 P^3 \left(\sqrt{3} \log ^{\frac{3}{2}}2\right)}+\cdots,
\end{eqnarray}
which allows a trivial real solution as, $ \eta_s \sim 0$.

For the range, $ \eta \geq 1 $, a similar analysis for $ +_{P, N_c} $ quivers (\ref{e32}) reveals an equation 
\begin{eqnarray}
\left(\eta_s ^2-1\right) \log \left(\frac{\eta_s +1}{\eta_s -1}\right)+2 \eta_s =0,
\end{eqnarray}
which can be solved using the method of transcendental equations.

Finally, considering fluctuations $ \delta\chi \sim y(\tau) $ about this invariant plane $ \lbrace p_{\chi}=0, \chi =0 \rbrace$, and retaining ourselves only upto leading order one finds simple harmonic motion (with a\emph{ unique} frequency $ \varpi \sim \ell $) of the following form
\begin{eqnarray}
\label{e87}
\ddot{y}(\tau)+\varpi^{2}y(\tau)\approx 0,
\end{eqnarray}
which guarantees the trivial integrability of the type IIB string soliton under consideration. 

In fact, this result is quite reminiscent of what has been observed previously in the context of type IIB $ AdS_5 \times S^5 $ superstrings \cite{Basu:2011fw}. One therefore do not in fact need to go through all the details of Kovacic's algorithm those were mentioned previously as the power of differential Galois theory confirms the underlying integrable structure of (\ref{e87}).
\subsubsection{Case II}
We now look for a different possibility, where we try to figure out the NVEs considering an expansion about some fixed $ \lbrace p_{\eta},\eta\rbrace $ plane of the four dimensional phase space.

To start with, an obvious choice would be look for solutions subjected to the constraint $ p_{\eta} \sim \dot{\eta}=0$ . A natural choice that is consistent with this constraint also amounts to set $ \ddot{\eta} =0$. Finally, a straightforward computation reveals that, $ \partial_{\eta}f_i \Big |_{\eta =0}=0 $ for $ i=1,2,3,4 $. 

To summarise, we therefore conclude that $ \eta =\dot{\eta}=\ddot{\eta}=0 $ is a consistent choice to start with which also solves the $ \eta $- equation (\ref{e81}). Our goal would be to consider fluctuations about this $ \lbrace p_{\eta}=0, \eta =0\rbrace $ plane and obtain the corresponding NVE.

Substituting this ansatz into (\ref{e82}), we obtain the corresponding solution to the $ \chi $- equation in terms of Jacobi amplitudes
\begin{eqnarray}
\label{e91}
\chi(\tau)=\text{am}\left(\sqrt{\left(c_1+\ell^2\right) \left(c_2+\tau \right){}^2}|\frac{\ell^2}{c_1+\ell^2}\right),
\end{eqnarray}
where $ c_1 $ and $ c_2 $ are the integration constants.

Using (\ref{e91}), we finally arrive at the NVE corresponding to $ \tilde{T}_{N_c,P} $ quivers (\ref{e21})
\begin{eqnarray}
\label{E92}
\ddot{y}(\tau)+\mathcal{A}(\tau)y(\tau)\approx 0,
\end{eqnarray}
where we identify the coefficient as
\begin{eqnarray}
\mathcal{A}(\tau) = 2\ell^2 \sin^2\chi -c_1 -\frac{1}{4}.
\end{eqnarray}

The general solution of (\ref{E92}) is quite complicated due to the presence of the function $ \mathcal{A}(\tau) $, hence the corresponding solution is \emph{non-Liouvillian}. This particular string embedding therefore clearly indicates the signature of non-integrability in the system. 

Setting the integration constants to zero, and considering an expansion for small parameter range ($ \tau \ll 1 $) it is possible to simplify the function
\begin{eqnarray}
\mathcal{A}(\tau) & \sim & 2\ell^2 \tanh ^2\left(\sqrt{\ell^2 \tau ^2}\right) -\frac{1}{4}\nonumber\\
& \sim & 2\ell^4 \tau^2  -\frac{1}{4},
\end{eqnarray}
which leads to solutions in terms of special parabolic cylindrical functions of the form
\begin{eqnarray}
\label{E95}
y(\tau)=c_1 D_{-\frac{1}{2}+\frac{i}{8 \sqrt{2} \ell^2}}\left((1+i) \sqrt[4]{2} \ell \tau \right)+c_2 D_{-\frac{1}{2}-\frac{i}{8 \sqrt{2} \ell^2}}\left((-2)^{3/4} \ell \tau \right).
\end{eqnarray}

Looking back at the Schroedinger form (\ref{e74}), we notice that the potential function $ V(\tau)=-\mathcal{A}(\tau) $ which makes the exact solution of $ \omega(\tau) $ quite difficult. In fact, for the small parameter range $ \tau \ll 1 $, the solution manifests itself in the form of special parabolic cylindrical functions those were mentioned in (\ref{E95}). This further confirms the non-Liouvillian nature of solutions for the NVE (\ref{E92}).

To summarize, our analysis reveals the existence of at least one particular phase space configuration that does not meet the Liouvillian integrability criteria those were mentioned previously. Based on the notion of holography, this naturally leads us to conjecture about the non-integrability of $ \mathcal{N}=1 $  linear quivers in $ 5d $. 
\section{Numerics}
\label{S4}
We now aim to decode the signatures of the above non-integrability in terms of various physical phenomena associated with the dynamical phase space under consideration. One natural quest along this direction would be to search for possibilities of a \emph{chaotic} motion for these type IIB strings by computing appropriate chaos indicators. 

For the purpose of our present analysis, we look for two such indicators namely (i) the Lyapunov exponent and (ii) the Poincare section. Below, we elaborate on each of these entities in detail. Finally, it is noteworthy to mention that the non-integrability does not necessarily imply a chaotic motion, although the reverse is always true.
\subsection{Chaos}
The purpose of this Section is to complement our analytical finding through numerics. In particular, we estimate the Lyapunov exponents ($ \lambda $) for the type IIB strings under consideration. This amounts of solving the set of equations (\ref{e81})-(\ref{e82}) for a given choice of the initial conditions. These initial conditions are set in such a manner so that the Hamiltonian constraint (\ref{E81}) vanishes identically. This naturally identifies the corresponding energy ($ E $) of the string soliton and/or the long quiver. For technical simplicity, we explore these exponents for $ \tilde{T}_{N_c,P} $ quivers (\ref{e21}) only.

\begin{figure}
\includegraphics[scale=.55]{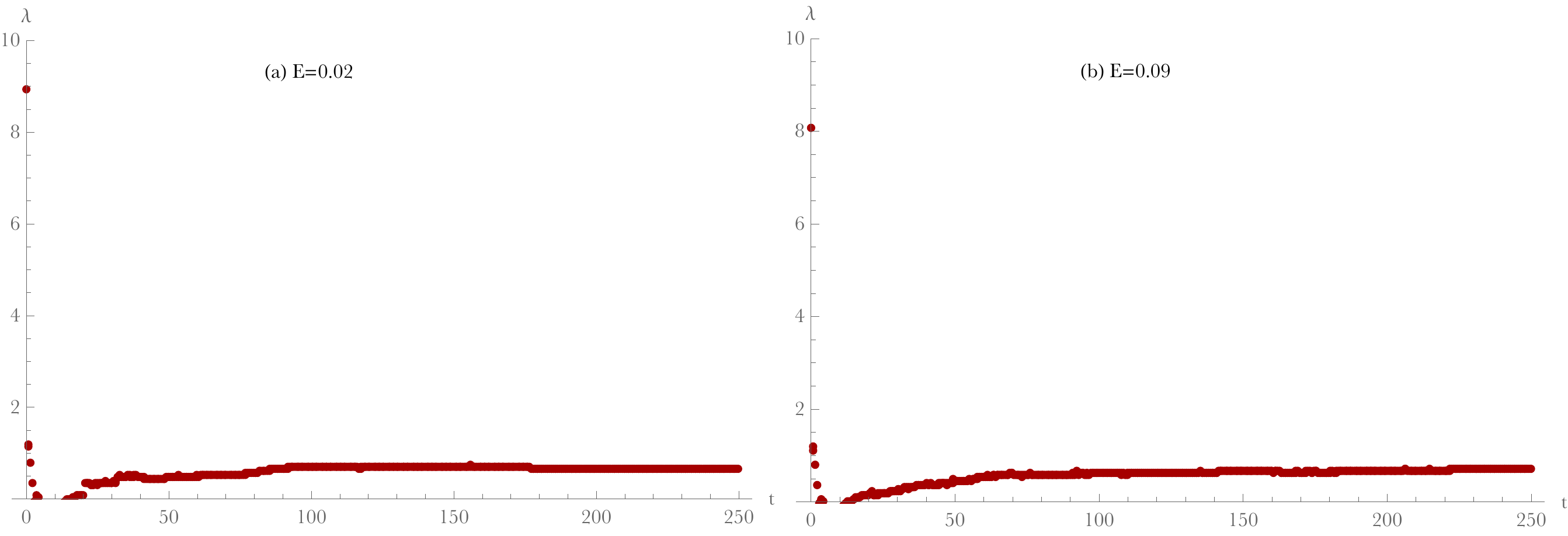}
  \caption{We plot the Lyapunov exponents ($ \lambda $) for $ \tilde{T}_{N_c,P} $ quivers. We set $ P=100 $ and $ \ell =5 $ for each of these plots. (a) Single kink profile that corresponds to a low energy quiver, (b) Single kink profile that corresponds to a higher excited state of the quiver. In the dual string theory picture, these plots correspond to strings those are less energetic and are located far away from the flavour D7 branes.}\label{lypnv}
\end{figure}

The non zero Lyapunov exponent, in some sense, is a measure of the \emph{chaotic} dynamics associated with a Hamiltonian system \cite{Zayas:2010fs}-\cite{Basu:2012ae}
\begin{eqnarray}
\label{e90}
\lambda = \lim_{\tau \rightarrow \infty}\lim_{\Delta X_0 \rightarrow 0}\frac{1}{\tau}\log \frac{\Delta X (X_{0}, \tau)}{\Delta X (X_0 ,0)}.
\end{eqnarray}

Typically, (\ref{e90}) serves the purpose of a quantitative measurement of the rate of separation between two nearby trajectories (in the phase space) for a small change in the initial conditions. Here, $ X_0 $ corresponds to the initial phase space data while the function $ \Delta X (X_0 , \tau)$ measures the separation between two infinitesimally close trajectories at sufficiently late times, for a small change in the initial conditions. 

Notice that, since the phase space of the present solitonic configuration is four dimensional $ \lbrace \eta, \chi, p_{\eta}, p_{\chi}\rbrace $ therefore in principle there exist four Lyapunov exponents for the system whose sum vanishes to zero namely $ \sum_{i=1}^{4}\lambda_i=0 $. However, for the present analysis, we look for the largest possible Lyapunov associated with the solitonic configuration which is sufficient to convince about the chaotic dynamics of the phase space. 

To begin with, we place the string soliton near the north pole of $ S^2 $ (as this naturally reduces the energy of the string) and explore the time evolution of the system. The initial conditions for Fig.\ref{lypnv}a correspond to setting, $ \chi (0)=0.05,\eta (0)=0.01,\dot{\chi}(0)=0.01 $ and $  \dot{\eta}(0)=0.001 $. These initial conditions suggest that we place the string solitons far away from the flavour D7 branes. This fixes the energy of the string soliton to be $ E \sim 0.02$ such that the constraint (\ref{e84}) is identically satisfied. From Fig.\ref{lypnv}a, it is quite evident that the Lypunov ($ \lambda $) asymptotically approaches to some non zero value. This clearly signifies the onset of chaos and hence non-integrability associated with the solitonic configuration.

We further excite these strings (Fig.\ref{lypnv}b) by placing them slightly away from the north pole. The initial conditions in this case correspond to setting $ \chi (0)=0.2,\eta (0)=0.01,\dot{\chi}(0)=0.01 $ and $  \dot{\eta}(0)=0.001 $. This fixes the energy of the string $ E \sim 0.09 $ which suggests that these solitons posses larger energy than those previous ones. Like before, we observe a non zero value of the Lyapunov exponent ($ \lambda $) which signals the persistence of chaotic motion associated with these type IIB strings.

\begin{figure}
\includegraphics[scale=.55]{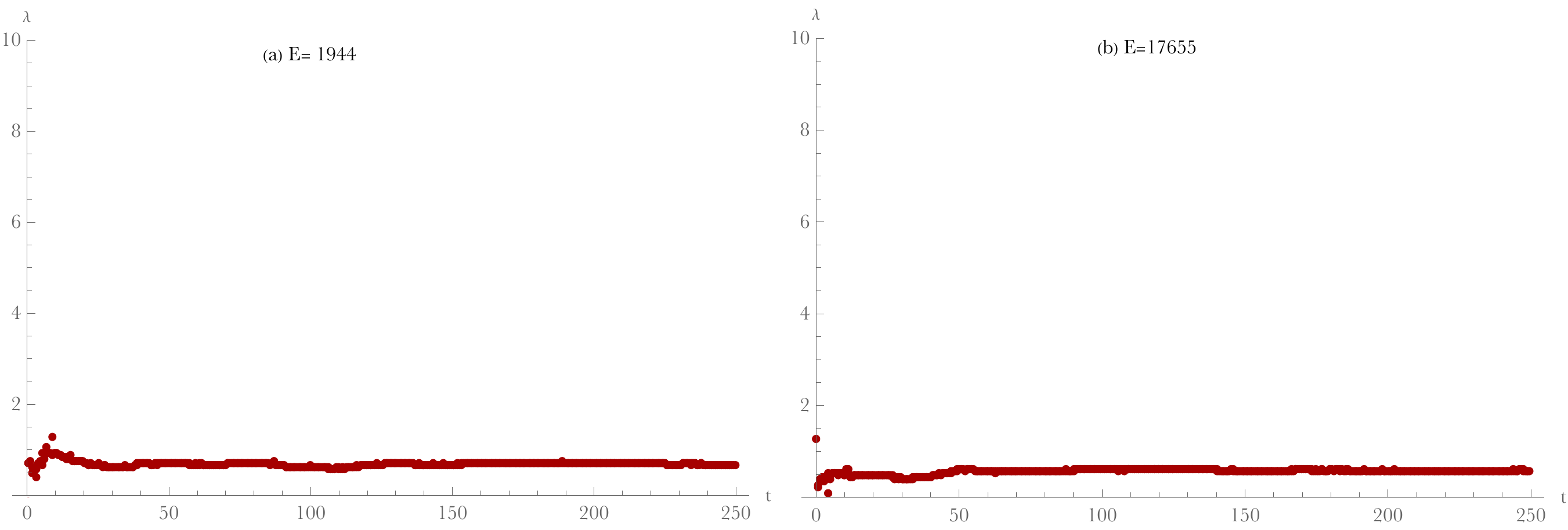}
  \caption{We plot the Lyapunov exponents ($ \lambda $) for $ \tilde{T}_{N_c,P} $ quivers. We set $ P=100 $ and $ \ell =5 $ for each of these plots. (a) Single kink profile that corresponds to a quiver with lower energy, (b) Single kink profile that corresponds to a higher excited state of the quiver. In the dual string theory picture, these plots correspond to highly energetic strings those are located nearer to the flavour D7 branes.}\label{Lypnv}
\end{figure}

Let us now consider a situation in which strings are initially placed closer to the flavour D7 branes. This is achieved by choosing an appropriate initial condition for the holographic coordinate $ \eta $. In the Fig\ref{Lypnv}. we show the corresponding plots for the Lyapunov exponents as the string approaches closer to the location of the flavour D7 branes which for the present example corresponds to, $ \eta \sim 100 $. From Fig \ref{Lypnv}. it is evident that the string becomes more and more energetic as it approaches the flavour branes. In other words, more energy needs to be supplied while moving the string closer to the location of the flavour branes. For our case, the initial positions for the string are set as $ \eta (0)=15 $ (Fig.\ref{Lypnv}a) and $ \eta (0)=75 $ (Fig.\ref{Lypnv}b) which correspond $ E \sim 1944 $ and $ E \sim 17655 $ respectively.
\subsection{Poincare section}
Usually, for an integrable system the phase space trajectories lie on the KAM torus \cite{Zayas:2010fs}. Generally, for an integrable system with $ N $ conserved charges ($ \mathcal{Q}_i $) - the corresponding KAM torus is $ N $ dimensional. The trajectories over the KAM torus are completely specified in terms of these $ N $ conserved charges ($ \mathcal{Q}_i $).

One elegant way to check the underlying integrable structure of the dynamical phase space is to take a $ 2d $ circular cross-section of these KAM tori and see whether a large number of \emph{foliated} circular KAM curves survive the external perturbation applied to the Hamiltonian of the system. These $ 2d $ cross-sections of the foliated KAM tori are known as the \emph{Poincare} section. As per the KAM theorem, any non-integrable perturbation added to the original integrable Hamiltonian destroys some of these KAM tori and thereby the foliated circular KAM curves those span the associated $ 2d $ Poincare sections. As the strength of the non-integrable deformation increases further, more and more KAM tori will be destroyed which will result in a seemingly random motion in the phase space.

\begin{figure}
\includegraphics[scale=.59]{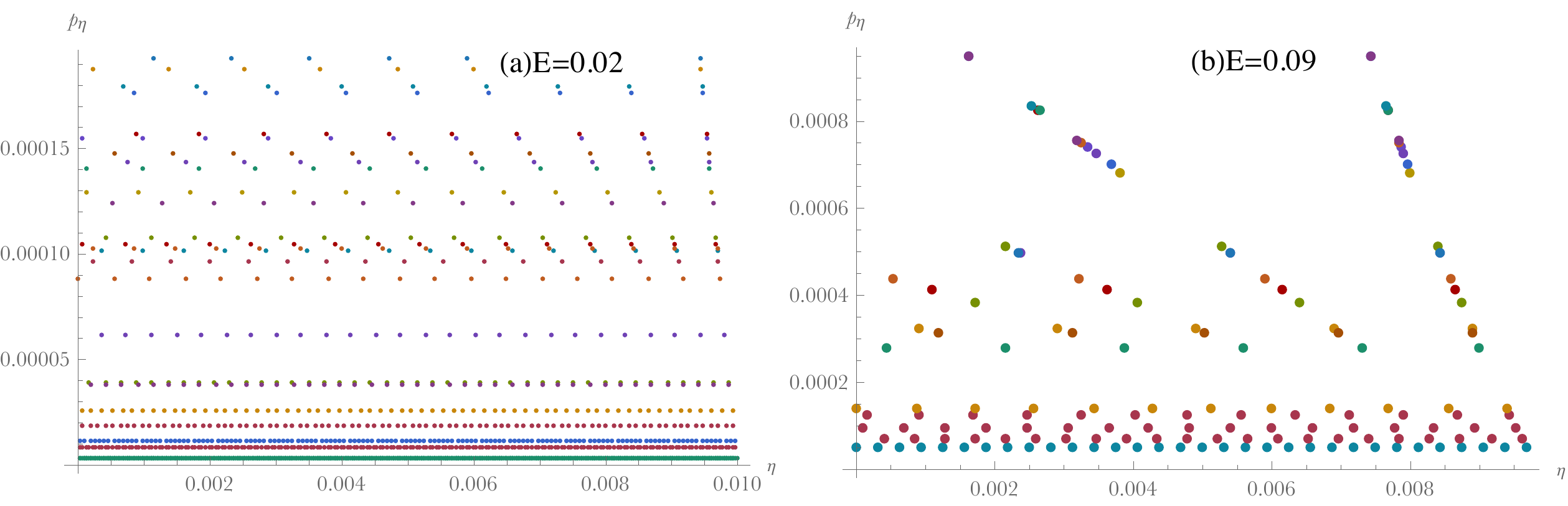}
  \caption{We plot the Poincare section for $ \tilde{T}_{N_c,P} $ quivers. We set $ P=100 $ and $ \ell =5 $ for each of these plots. (a) Single kink profile that corresponds to a low energy quiver, (b) Single kink profile that corresponds to a higher excited state of the quiver. In the dual string theory picture, these plots correspond to a phase space configuration with less energetic strings those are located far away from the flavour D7 branes.}\label{poincare}
\end{figure}

In order to probe these Poincare sections corresponding to $ \mathcal{N}=1 $ quivers in $ 5d $, we solve the corresponding Hamiltonian dynamics that results from (\ref{e81}). Like before, we set the energy ($ E =E_0 $) of the string at some particular value which satisfies the Hamiltonian constraint (\ref{e84}) for some given set of initial conditions namely, $ \eta(0)=0.01 $ and $ p_{\chi}(0)=0 $. Given these initial conditions, we generate a random data set by choosing $ p_{\eta}\in [0 , 10] $ which fixes the corresponding $ \chi(0) $ such that the constraint (\ref{e84}) is always satisfied\footnote{Following our previous discussion, we stick to a configuration where the energy ($ E $) of the soliton is the lowest possible. This corresponds to strings located away from the flavour D7 branes. For our purpose, it is sufficient to explore the Poincare section for these low energy strings since the increase in $ E $ would enhance the possibilities of a more random motion which will destroy the associated KAM tori in the phase space. }.

Finally, with the help of this initial data set, we carry out a numerical simulation of the Hamilton's equations of motion
\begin{eqnarray}
\dot{\chi}&=&-\frac{1}{2f_2}(p_{\chi}-2 \ell f_4 \sin\chi),\\
\dot{p}_{\chi}&=&\ell^2 \left(f_2 +\frac{f^2_4}{f_2} \right) \sin 2\chi -\frac{\ell f_4}{f_2}p_{\chi}\cos\chi ,\\
\dot{\eta}&=&-\frac{p_{\eta}}{2f_3},
\end{eqnarray}
\begin{eqnarray}
\label{e100}
\dot{p}_{\eta}=\frac{E^2}{4f^2_1}\partial_{\eta}f_1 -\frac{p^2_{\eta}}{4 f^2_3}\partial_{\eta}f_3 -\frac{\partial_{\eta}f_2}{4f^2_2}(p_{\chi}-2\ell f_4 \sin\chi)^2\nonumber\\
-\frac{\ell}{f_2}\partial_{\eta}f_4 \sin\chi (p_{\chi}-2\ell f_4 \sin\chi)+\ell^2 \partial_{\eta}f_2 \sin^2\chi ,
\end{eqnarray}
and plot all the points on the $ \lbrace  p_{\eta} , \eta \rbrace $ plane every time the trajectories pass through $ \chi (t) =0 $ hyper-plane. Which therefore represents a two dimensional projection of a three dimensional hyper-plane in the phase space.

For an integrable phase space configuration, a naive expectation would be to see the patches of circular KAM curves through this $ 2d $ projection. For the present analysis, the phase under consideration is four dimensional namely it is characterized by the axes $ \lbrace \chi , p_{\chi},\eta , p_{\eta}\rbrace $. In case of an integrable phase space, one would therefore expect the trajectories to be aligned along the two dimensional torus. Poincare section of this torus would therefore unveil circular patches indicating the presence of different resonant tori.

From above Fig.\ref{poincare}, however, we do not see any evidence for such closed patches. As mentioned above, these Poincare sections are obtained for strings sitting at the north pole ($ \chi =0 $). As our analysis reveals, at low enough energies the trajectories in the phase space are more ``organized" than its high energy counterparts. Most of the string orbits are \emph{localized} near the north pole of $ S^2 $ while the strings can move along the holographic direction ($ \eta $) (Fig.\ref{poincare}a). However, as the energy increases, the string starts moving with higher momentum ($ p_{\eta} $) which by virtue of (\ref{e100}) reveals that the solitonic orbits are no longer localized near the north pole of $ S^2 $. In other words, the string starts moving randomly along the two sphere which results in a chaotic distribution of points along the Poincare section (Fig.\ref{poincare}a). The distribution of the these points (along Poincare section) become more and more sparse as the corresponding energy ($ E $) of the soliton increases. 

To summarise, both these analyses accumulate enough evidence for the underlying non-integrable structure of the $ \mathcal{N}=1 $ SCFTs in $ 5d $.
\section{Summary and final remarks}
\label{S5}
To conclude, the present analysis reveals the existence of yet another class of strongly coupled \emph{non-integrable} SCFTs among the plethora of examples those are floating in the AdS/CFT landscape. These are the $ \mathcal{N}=1 $ SCFTs living in $ 5d $ whose dual stringy counter part is described by type IIB supergravity solutions in $ AdS_6 \times S^2 \times \Sigma_{(2)}$. 

Our analysis reveals that these $ 5d $ theories are Liouvillian non-integrable in the sense of \cite{kov1}-\cite{kov2}. We further complement our claim through numerics where we estimate the corresponding Lyapunov exponent as well as the Poincare section for the dynamical phase space under consideration. As a future remark, it is worthwhile to mention that it will be really nice to consider an S dual of these type IIB solutions andcheck whether the Liouvillian non-integrability of the transformed background is still preserved. In fact, as a first step, one should construct the corresponding S dual background and perform an analysis of various field theory observables for example - the Page charges as well as the central charges corresponding to different $ \mathcal{N}=1 $ quivers. 

It will be also interesting to explore whether it is possible to implement the algorithm for a more generic class of quivers in $ 5d $. These are the quivers with unbalanced nodes - for example, $ X_{N_c , M} $ theories as discussed in \cite{Uhlemann:2020bek} and the ($ Y_{N_c} $) quivers which are not S dual to themselves. In principle, the algorithm should be also be applicable for these theories based on the notion of  the $ SL(2, C) $ invariance of the space of solutions for complex PDEs those emerge from studying the dynamics of the string soliton over complex manifold.

We hope to be able to address some of these issues in the near future.\\

{\bf {Acknowledgements :}}
  The author is indebted to the authorities of IIT Roorkee for their unconditional support towards researches in basic sciences. The author is thankful to Carlos Nunez and Christoph Uhlemann for their valuable as well as thoughtful comments on the draft. The author would like to acknowledge The Royal Society, UK for financial assistance. The author would also like to acknowledge the Grant (No. SRG/2020/000088) received from The Science and Engineering Research Board (SERB), India. \\ 
\appendix
\numberwithin{equation}{section}
\renewcommand{\theequation}{\thesection\arabic{equation}}
\section{Non-integrability of long extended strings}
\label{A}
Let us briefly discuss the Liouvillian non-integrability for long strings passing through flavour D7 branes located along the internal manifold ($ \mathcal{M}_4 $) of the full type IIB solution. The dynamics of these strings are described in (\ref{e17})-(\ref{e18}).

Clearly, a choice $ \sigma'' = \sigma' =\sigma=0 $ solves (\ref{e17}) which we therefore choose to be the reference plane about which the normal fluctuations ($ \delta\sigma \sim y(\tilde{\sigma}) $) are considered.

Substituting this choice into (\ref{e18}), we find
\begin{eqnarray}
\label{a1}
\eta'' \approx \left( \ell^2 - \frac{1}{4}\right) \eta + \cdots,
\end{eqnarray}
where we stick to the large $ P $ limit while taking the specific example of $ \tilde{T}_{N_c,P} $ quivers (\ref{e21}).

The above equation (\ref{a1}) is solved for
\begin{eqnarray}
\label{a2}
\eta_{s} (\tilde{\sigma}) \simeq e^{-\gamma \tilde{\sigma}}+\mathcal{O}(1/P^2),
\end{eqnarray}
where, $ \gamma = \sqrt{ \ell^2 - \frac{1}{4}} $.

Using (\ref{a2}), the corresponding NVE turns out to be
\begin{eqnarray}
y''(\tilde{\sigma})+\mathcal{B}(\tilde{\sigma})y'(\tilde{\sigma})+\mathcal{A}(\tilde{\sigma})y(\tilde{\sigma})\approx 0,
\end{eqnarray}
where the individual coefficients are identified as
\begin{eqnarray}
\label{a3}
\mathcal{B}(\tilde{\sigma}) &=&\frac{\pi^2 \eta_s \eta'_s}{24 P^2 \log 2 },\\
\mathcal{A}(\tilde{\sigma}) &=&-\left( \frac{1}{2}+\frac{\pi^2 (\ell^2 \eta^2_s +\eta'^2_s)}{24 P^2 \log 2 }\right). 
\end{eqnarray}
Clearly, the general solution of (\ref{a3}) is non-Liouvillian. However, if we retain ourselves to the strict supergravity ($ P \rightarrow \infty $) limit then the solution is of course Liouvillian, $ y(\tilde{\sigma})\sim e^{\frac{\tilde{\sigma}}{\sqrt{2}}} $. This is the limit in which the dispersion relation(s) (\ref{e30}) has been computed.

As a special note, below we express the potential function associated with the corresponding Schrodinger form
\begin{eqnarray}
-V(\tilde{\sigma})\simeq -\frac{1}{2}+\frac{\pi^2 \ell^2}{24 P^2 \log2}e^{-2\gamma \tilde{\sigma}}+\mathcal{O}(1/P^4),
\end{eqnarray}
which reveals a solution of (\ref{e74}) in terms of special functions namely a combination of Bessel functions coupled together with the Gamma functions. On the other hand, in the strict holographic limit ($ P \rightarrow \infty $), the solution ($ \omega (\tilde{\sigma}) $) corresponding to the Schrodinger equation (\ref{e74}) manifests itself as a polynomial of degree 1, which confirms the Liouvillian integrability of the (long) string soliton in the above limit.

To summarize, we notice that it is the presence of flavor D7 branes which spoils the integrability. Setting, $ P \rightarrow \infty $ corresponds to the fact that we place the flavour branes at infinity - as a consequence of this the string never meets the flavor branes. This is the limit in which the dispersion relation(s) (\ref{e30}) has been obtained. These are the massive string states which in fact cannot be excited.
\section{A note on the large $\sigma $ limit}
\label{B}
Below, we enumerate the metric functions ($ f_i (\sigma , \eta) $) and their derivatives in the large $ \sigma $ limit. Given the potential function as in (\ref{e23}), the corresponding metric functions read as,
\begin{eqnarray}
\label{A1}
f_1(\sigma , \eta) &\sim & \frac{3 \sqrt{3} P}{2}+\frac{1}{4} \sqrt{3} 3 \pi  \sigma +\mathcal{O}(1/P),\\
f_2(\sigma , \eta) &\sim &\frac{\sqrt{3} \pi ^2 \eta ^2}{2 P}+\mathcal{O}(1/P^2),\\
f_3(\sigma , \eta) &\sim &\frac{\sqrt{3} \pi ^2}{2 P}+\mathcal{O}(1/P^2),\\
f_4(\sigma , \eta) &\sim &\frac{4 \pi ^3 \eta ^3}{3 P^2}+\mathcal{O}(1/P^3),\\
f_5(\sigma , \eta) &\sim & -\frac{3 \left(\pi ^3 \eta ^3 N_c\right)}{2 P^2} +\mathcal{O}(1/P^3),\\
f_6(\sigma , \eta) &\sim & \frac{4 N_c^2}{3}+\mathcal{O}(1/P),\\
f_7(\sigma , \eta) &\sim &-2 N_c +\mathcal{O}(1/P).
\label{A7}
\end{eqnarray}

Using the above metric forms (\ref{A1})-(\ref{A7}), we find
\begin{eqnarray}
\label{A8}
 \ddot{\eta}& \approx & \eta (\dot{\chi}^2 -\ell^2 \sin^2\chi),\\
 -\eta \ddot{\chi}& \approx &2 \dot{\eta}\dot{\chi}+\ell^2 \eta \sin\chi \cos\chi.
 \label{A9}
\end{eqnarray}

In order to check the Liouvillian non-integrability criteria, we set $ \ddot{\chi}=\dot{\chi}=\chi=0 $ which clearly solves (\ref{A9}). In other words, we choose to work with an invariant plane $ \lbrace p_{\chi}=0, \chi =0\rbrace $ in the phase space and thereafter consider fluctuations about this plane.

Given the above choice, from (\ref{A8}) we find
\begin{eqnarray}
\label{A10}
\eta (\tau) |_{\chi \sim 0}\sim a \tau + b.
\end{eqnarray}

Substituting (\ref{A10}) into (\ref{A9}) and considering fluctuations $ \delta \chi \sim y(\tau) $, the corresponding NVE turns out to be
\begin{eqnarray}
\ddot{y}(\tau)+\mathcal{B}(\tau)\dot{y}(\tau)+\mathcal{A}(\tau)y(\tau)\approx 0,
\end{eqnarray}
where we define coefficients as, $ \mathcal{B}(\tau)=\frac{2a}{a \tau +b} $ and $ \mathcal{A}(\tau)=\ell^2 <0$.

The corresponding solution turns out to be Liouvillian namely
\begin{eqnarray}
y(\tau)\sim \frac{e^{-\sqrt{-\ell^2} \tau } \left(\frac{c_2 e^{2 \sqrt{-\ell^2} \tau }}{\sqrt{-\ell^2}}+2 c_1\right)}{2 (a \tau +b)}~;~| \tau | \ll 1.
\end{eqnarray}

The potential function for the Schrodinger equation (\ref{e74}) turns out to be
\begin{eqnarray}
V(\tau)=- \ell^2 ,
\end{eqnarray}
which is quite analogous to the case of simple harmonic motion as discussed in Section \ref{3.3.1} which for small enough fluctuations $ | y(\tau) | \ll 1 $ yields a solution of the form
\begin{eqnarray}
\omega (\tau)\sim -\ell^2 \tau
\end{eqnarray}
which is a polynomial of degree 1.

Let us now look at the other possibility namely to set, $\ddot{\eta}=\dot{\eta}=\eta =0  $ which clearly solves (\ref{A8}). Substituting this choice into (\ref{A9}) we find
\begin{eqnarray}
\ddot{\chi}+2 g(\tau)\dot{\chi}+\ell^2 \sin\chi \cos\chi \approx 0,
\label{A15}
\end{eqnarray}
where we introduce, $ \lim_{\eta \rightarrow 0}\frac{\dot{\eta}(\tau)}{\eta (\tau)} = g(\tau)$. 

Naturally, any possible solution of (\ref{A15}) is quite nontrivial which when substituted back into the NVE corresponding to fluctuations ($ \delta\eta \sim y(\tau) $) about the hyper-plane $ \lbrace p_{\eta}=0 , \eta =0\rbrace $ yields solutions which are non-Liouvillian in nature.
\section{Spacetime solutions ($ f_i(\sigma , \eta) $) for $ \eta> 1 $}
\label{C}
Below we summarize metric components for the range $ 1\leq \eta \leq P $. We first note down
\begin{eqnarray}
 f_1(\sigma \sim 0, \eta) \sim \frac{3\sqrt{3}}{2\sqrt{2}}  \pi  \sqrt{\frac{a_1(\eta)}{b_1(\eta)}}
\end{eqnarray}
where we denote the individual entities as
\begin{eqnarray}
a_1 (\eta)&=&-\left(\eta ^2+1\right) \log \left(\frac{\eta +1}{\eta -1}\right)+\eta  \left(-2 \log \left(\eta ^2-1\right)+6+\log 16\right)-4 \eta  \log \left(\frac{\pi }{P}\right),\\
b_1(\eta)&=&\log \left(\frac{\eta +1}{\eta -1}\right).
\end{eqnarray}

Next, we note down the function $ f_2 $ which can be schematically expressed 
\begin{eqnarray}
 f_2(\sigma \sim 0, \eta) \sim \frac{a_2(\eta)}{b_2(\eta)}
\end{eqnarray}
in terms of other functions
\begin{eqnarray}
\frac{a_2 (\eta)}{\sqrt{3} \pi}&=&  \left(\left(\eta ^2+1\right) \log \left(\frac{\eta +1}{\eta -1}\right)-2 \eta  \left(-\log \left(\eta ^2-1\right)+3+\log (4)\right)+4 \eta  \log \left(\frac{\pi }{P}\right)\right)^2,\\
b_2(\eta)&=&c_2(\eta)\sqrt{\frac{-2 \left(\eta ^2+1\right) \log \left(\frac{\eta +1}{\eta -1}\right)+4 \eta  \left(-\log \left(\eta ^2-1\right)+3+\log (4)\right)-8 \eta  \log \left(\frac{\pi }{P}\right)}{\log \left(\frac{\eta +1}{\eta -1}\right)}},\\
c_2(\eta)&=&-2 \left(\eta ^2+3\right) \log ^2\left(\frac{\eta +1}{\eta -1}\right)+\left(-2 \log \left(\eta ^2-1\right)+4+\log (16)\right)^2+16 \log ^2\left(\frac{\pi }{P}\right).
\end{eqnarray}

Finally, we note down the metric component
\begin{eqnarray}
f_3 (\sigma \sim 0, \eta)\sim -\frac{a_3(\eta)}{b_3(\eta)}
\end{eqnarray}
where we identify individual entities as
\begin{eqnarray}
a_3(\eta)=\sqrt{\frac{3}{2}} \pi  \log \left(\frac{\eta +1}{\eta -1}\right)~~~~~~~~~~~~~~~~~~~~~~~~~~~~~~~~~~~~~~~~~~~~~~~~~~~~~~~~~~~~~~~~~~~~\nonumber\\
\times \sqrt{\frac{-\left(\eta ^2+1\right) \log \left(\frac{\eta +1}{\eta -1}\right)+\eta  \left(-2 \log \left(\eta ^2-1\right)+6+\log (16)\right)-4 \eta  \log \left(\frac{\pi }{P}\right)}{\log \left(\frac{\eta +1}{\eta -1}\right)}},
\end{eqnarray}
\begin{eqnarray}
b_3(\eta)=\left(\eta ^2+1\right) \log \left(\frac{\eta +1}{\eta -1}\right)-2 \eta  \left(-\log \left(\eta ^2-1\right)+3+\log (4)\right)+4 \eta  \log \left(\frac{\pi }{P}\right).
\end{eqnarray}

\end{document}